\documentclass[reprint, 10pt, a4paper, aip, floatfix]{revtex4-1}
\usepackage[utf8]{inputenc}
\usepackage{amsmath}
\usepackage{amsfonts}
\usepackage{amssymb}
\usepackage{mathrsfs}
\usepackage{subfigure}
\usepackage[hidelinks]{hyperref}
\usepackage{outlines}
\usepackage{tikz}
\usepackage{booktabs}

\newcommand{\bx}{\mathbf{x}}

\newcommand{\bmu}{\ensuremath{\boldsymbol\mu}}

\newcommand{\WGmu}{\Delta W_{\text{G}|\mu}}
\newcommand{\WG}{\Delta W_\text{G}}
\newcommand{\WDmu}{\Delta W_{\text{D}|\mu}}
\newcommand{\WD}{\Delta W_\text{D}}

\begin{document}
	
\title{Available Energy from Diffusive and Reversible Phase Space Rearrangements}
\date{\today}
\author{E.~J. Kolmes}
\email[Electronic mail: ]{ekolmes@princeton.edu}
\affiliation{Department of Astrophysical Sciences, Princeton University, Princeton, New Jersey 08540, USA}
\author{P. Helander}
\affiliation{Max-Planck Institut f\"ur Plasmaphysik, 17491 Greifswald, Germany}
\author{N.~J.~Fisch}
\affiliation{Department of Astrophysical Sciences, Princeton University, Princeton, New Jersey 08540, USA}

\begin{abstract}
Rearranging the six-dimensional phase space of particles in plasma can release energy.
The rearrangement may happen through the application of electric and magnetic fields, subject to various constraints. 
The maximum  energy that can be released through a rearrangement of a distribution of particles can be called its available or free energy. 
Rearrangement subject to phase space volume conservation leads to the classic Gardner free energy. 
Less free energy is available when constraints are applied, such as respecting conserved quantities. 
Also, less energy is available if particles can only be diffused in phase-space  from regions of high phase-space density to regions of lower phase-space density. 
The least amount of free energy is available if particles can only be diffused in phase space, while conserved quantities still need to be respected. 
\end{abstract}
	
\maketitle
	
\section{Introduction}

Waves injected into plasma can be amplified, extracting energy from the plasma. 
Similarly, internal modes within the plasma can grow by extracting energy from the plasma.
There are multiple and distinct ways in which this energy can be accessed, and the open question is how much energy can possibly be accessed. 
The ground state of the system can be defined as the state of least energy accessible respecting any constraints. 
The free energy or the accessible energy can then be defined as the difference between the initial state energy and the ground state energy.

In his classic work,\cite{Gardner1963} Gardner calculated the free energy when plasma could be rearranged while preserving the six-dimensional phase space densities or volumes. 
If the distribution is divided into separate phase space volumes, the volume preservation constraint means that it is not possible to rearrange the system to any lower-energy state than the one in which the lowest-energy regions of phase space are occupied by the most-populated available phase space elements. 
Each volume carries with it its initial number of particles, so putting that volume in a six-dimensional location of lower energy preserves that volume phase space density as well. 
This is known as Gardner restacking, representing a particular phase space rearrangement.
The fully restacked phase space is the ground state subject to phase space conservation; the releasable energy is known as the Gardner free energy.

This problem was approached using variational techniques by Dodin and Fisch.\cite{Dodin2005}
Dodin and Fisch also calculated the free energy,  including the additional constraint that the total current be preserved.  
The free energy respecting current conservation, of course, would be less than the Gardner free energy.
Recently, Helander  calculated the free energy, conserving like Gardner the six-dimensional phase space densities, but conserving as well other quantities, particularly those quantities that affect the motion of individual particles in phase space.\cite{Helander2017ii, Helander2020} 
For example, in a magnetized system,  quantities like the first or second adiabatic invariants $\mu$ and $J$ might be conserved. 
The free energy constrained by conservation of the adiabatic invariants of the motion, of course, would similarly be less than the Gardner free energy.

However, even under Hamiltonian dynamics,  rearrangements of plasma can appear not to conserve phase space volume, when the distribution functions are viewed with  finite granularity. 
Viewed at finite granularity,  waves can diffuse particles.\cite{Kennel1966}  
Importantly, these diffusive wave-particle interactions can release particle energy to the waves.\cite{Fisch1992}
The  energy accessible via diffusive operations in phase space  is less than that accessible via restacking. 
The property of free energy constrained by phase space diffusion was first posed by Fisch and Rax.\cite{Fisch1993}

The operation of  diffusing particles between volumes in phase space, rather than  interchanging these volumes, has actually had  applicability to a rather wide range of problems.
Mathematical treatments of rearrangement by diffusion describe theories of income inequality,\cite{Dalton1920} altruism,\cite{Thon2004} and physical chemistry.\cite{Horn1964, Zylka1985}
The maximal extractable energy under diffusive phase space rearrangements in plasma was  addressed by Hay, Schiff, and Fisch. \cite{Hay2015, Hay2017}

Helander's recent calculation of the plasma free energy obeying phase space conservation, with the motion of individual particles constrained by adiabatic invariants, now points to a natural generalization: the free energy under  diffusion in phase space, but with the motion of individual particles similarly constrained by adiabatic invariants.
This paper will discuss that generalization. 
Lest this be thought to be an academic exercise, we note that it is realized, for instance, in the quasilinear theory of plasma waves and instabilities with frequencies below the cyclotron frequency. 
The available energies subject to both diffusion and adiabatic motion constraints should be more limited than the available energy subject only to either constraint. 

The paper is organized as follows: 
Section~\ref{sec:availableEnergies}  describes these different available energies. 
Section~\ref{sec:discreteModel} introduces a simple, discrete model which illustrates the different ways in which these energies can be extracted. 
Section~\ref{sec:inhomogeneousField} shows how that simple model can be modified  to describe a more concrete plasma system. 
Finally, Section~\ref{sec:discussion} discusses the context and implications of these ideas. 

\section{The Four Classes of Available Energy} \label{sec:availableEnergies}

Let $\WG$ denote the accessible energy in Gardner's restacking problem, in which phase space volume conservation is the only restriction. Let $\WGmu$ denote Helander's available energy for the version of this problem in which one or more conservation law constraints are included. Let $\WD$ denote the maximum extractable energy in the variant of Gardner's problem in which phase space elements must be diffusively averaged rather than being exchanged (``restacked"). Finally, let $\WDmu$ denote the maximum extractable energy for the diffusive problem with conservation laws. 

Here, a ``diffusive exchange" refers to an operation in which the populations $f_a$ and $f_b$ of two equal-volume regions of phase space are mixed so that both populations are $(f_a + f_b) / 2$ after the exchange. 
There is no requirement that these two elements be adjacent in phase space; in fact, microscopically local flows can give rise to apparently non-local diffusive processes.\cite{Fisch1993} 
This kind of exchange can describe any process that tends to equalize the populations of different regions of phase space, regardless of the details of the microscopic dynamics. 
This equalization could, for instance, be the result of quasilinear diffusion but could equally well result from more complicated transport processes in phase space involving finite-Kubo-number effects, L\'evy flights, etc. 

This leaves us with four distinct available energies. 
The condition for the plasma to be in a ground state is the same in the restacking and diffusion problems: for any pair of equal-volume phase space elements with different populations, the higher-population element must occupy a region of phase space with no more energy than the lower-population element. If there is a conserved quantity $\mu$ (or, in general, a vector of conserved quantities $\bmu$), then the condition is identical except that it is only necessary to consider pairs of phase space elements with the same values of $\mu$ (or $\bmu$). 

Although Gardner restacking, even with constraints, leads to a unique ground state, that is not the case with diffusive exchange.
It is possible through diffusive exchange to reach more than one possible ground state from the same initial configuration. 
These ground states will often have different energies. 
Note that  $\WD$ and $\WDmu$ are defined as the maximum possible extractable energy, 
which is the difference between the initial energy and the energy of the lowest-energy accessible ground state. 

Given the same initial phase space configuration, diffusive exchange and restacking will typically not lead to the same ground state. The extractable energy will differ too.
It is clear that the available energies without additional conservation-law restrictions will always be at least as large as their restricted counterparts. That is, 
\begin{gather}
\WGmu \leq \WG \\
\WDmu \leq \WD \, . 
\end{gather}
Moreover, phase space restacking can always access at least as much energy as diffusive exchanges can (strictly more, if they are nonzero). To see this, note that both kinds of exchange move the system toward the same ground state condition, where the most highly populated phase space volumes are at the lowest energies, but that diffusive exchanges reduce the difference between the more- and less-highly populated phase space volumes. This reasoning is equally applicable with and without conservation laws, so it follows that 
\begin{gather}
\WD \leq \WG \\
\WDmu \leq \WGmu \, , 
\end{gather}
with equality holding only when both sides vanish. 

\section{Simple Discrete Model} \label{sec:discreteModel}

In order to get an intuitive sense for the four available energies -- that is, the restacking energy with and without a conservation law and the diffusive-exchange energy with and without a conservation law -- it is helpful to construct a simple model that shows concretely how the different energies can play out. 

\begin{figure}
	\includegraphics[width=\linewidth]{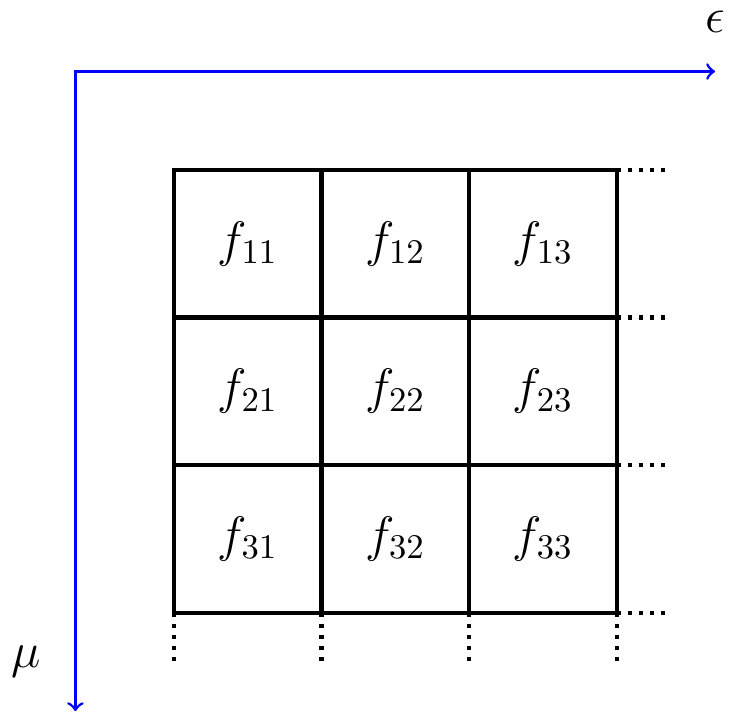}
	\caption{Schematic of a simple discrete system, with varying energy and some additional coordinate $\mu$. } 
	\label{fig:grid}
\end{figure}

To that end, consider a collection of discrete states, indexed by their energies and by some coordinate $\mu$, as shown schematically in Figure~\ref{fig:grid}. Associate the states in the $n$th column with energy $\epsilon_n$, with $\epsilon_i \leq \epsilon_j$ $\forall i < j$. One could construct a grid of this kind with any number of columns and rows. Physically, previous work presents the discrete version of these problems in two ways.\cite{Fisch1993, Hay2015, Hay2017} First, it can model an intrinsically discrete physical system, like transitions between atomic energy levels stimulated by lasers. Second, it can model a system with continuous phase space (e.g., a plasma) being mixed with some finite granularity. 

In this scenario, it is straightforward to understand the four available energies described in Section~\ref{sec:availableEnergies}. If $\mu$ is conserved, different rows are not allowed to interact, and the total available energy (either $\WGmu$ or $\WDmu$) is the sum of the available energies of the individual rows, considered independently. If $\mu$ is not conserved, then either the diffusive or the Gardner energy minimization problem must consider the system as a whole. 

When constructing examples of this kind, it quickly becomes apparent that conservation laws affect the outcome of diffusive and Gardner relaxation in qualitatively different ways. Consider the following phase-space configuration, with two energy levels (with the left-hand column corresponding to a lower energy) and two different values of $\mu$:
\begin{gather*}
\begin{array}{|c|c|}
\hline
0 & 1 \\ \hline
0 & 0 \\ \hline
\end{array} \; .
\end{gather*}
Suppose the separation between the two energy levels is $\varepsilon$. Then $\WGmu = \WG = \varepsilon$ and $\WDmu = \varepsilon / 2$. If $\mu$ is not conserved,, the lowest diffusively accessible energy can be reached by the sequence 
\begin{gather}
\begin{array}{|c|c|}
\hline
0 & 1 \\ \hline
0 & 0 \\ \hline
\end{array} \rightarrow 
\begin{array}{|c|c|}
\hline
1/2 & 1/2 \\ \hline
0 & 0 \\ \hline
\end{array} \rightarrow 
\begin{array}{|c|c|}
\hline
1/2 & 1/4 \\ \hline
1/4 & 0 \\ \hline
\end{array}
\end{gather}
so that $\WG = (3/4) \varepsilon$. At this point, we note that there is a qualitative difference between discrete and continuous systems. In the example just given, the ground state contradicts a theorem by Gardner that the distribution function can only depend on energy alone. This contradiction is a consequence of the fact that energetically neutral exchanges are possible between the two low-energy boxes in the diagram. It disappears if the distribution function is required to be a smooth function of a continuous energy variable.  

This example worked because, in the absence of any $\mu$ constraint, diffusive processes can take advantage of the additional phase space to more efficiently transfer material (or quanta, etc., depending on how this model is interpreted physically) more efficiently from high-energy to low-energy states. One might imagine that this a special property of phase space that is initially unoccupied. However, similar behavior appears when the initial population configurations for each $\mu$ are the same. Consider, for example, the following $2 \times 2$ system: 
\begin{align*}
\begin{array}{|c|c|}
\hline 
0 & 1 \\ \hline
0 & 1 \\ \hline
\end{array} \; .
\end{align*}
Suppose, as before, that the rows correspond to different values of $\mu$ and that the columns correspond to energy states that are separated by some energy $\varepsilon$. Three of the four measures of available energy are immediately clear: $\WGmu = \WG = 2 \varepsilon$ and $\WDmu = \varepsilon$. 

Hay, Schiff, and Fisch showed\cite{Hay2015} that there are only a finite number of candidates for the optimal sequence of diffusive steps to relax a system to equilibrium. After enumerating all possible candidates, it is possible to show that the lowest possible energy state accessible through diffusive steps can be reached by the sequence 
\begin{align*}
&\begin{array}{|c|c|}
\hline 
0 & 1 \\ \hline
0 & 1 \\ \hline
\end{array} \rightarrow
\begin{array}{|c|c|}
\hline 
0 & 1 \\ \hline
1/2 & 1/2 \\ \hline
\end{array} \rightarrow 
\begin{array}{|c|c|}
\hline 
0 & 3/4 \\ \hline
3/4 & 1/2 \\ \hline
\end{array} \\
&\hspace{120 pt} \rightarrow 
\begin{array}{|c|c|}
\hline 
1/4 & 3/4 \\ \hline
3/4 & 1/4 \\ \hline
\end{array} \rightarrow 
\begin{array}{|c|c|}
\hline 
1/2 & 1/2 \\ \hline
3/4 & 1/4 \\ \hline
\end{array} \; , 
\end{align*}
so $\WD = (5/4) \varepsilon$. 

It seems that there is some qualitative way in which increasing the number of accessible states improves the diffusive available energy without necessarily improving the Gardner available energy. However, it is nontrivial to write down a strong condition relating $\WGmu$, $\WG$, $\WDmu$, and $\WD$ that captures this intuition. 

One might expect, after looking at a number of examples involving small systems, that $\WDmu / \WD$ would always be less than or equal to $\WGmu / \WG$ (in other words, that restricting the size of accessible phase space would always have a fractionally more severe impact on the energy accessible via diffusive exchange). In fact, this inequality holds for any system with two energy levels (a $2 \times N$ grid of states). 

To see this, first note that in a $2 \times N$ system, $\WDmu = (1/2) \WGmu$. If only two cells can be exchanged, then it is either favorable to exchange them (in which case the Gardner exchange moves the difference between the cells' populations from one to the other and the diffusive exchange accomplishes half that) or it is not (in which case neither does anything). 

Now consider the same system without any $\mu$ constraints. It is possible to divide the initial population values $\{f_{ij}\}$ into the $N$ largest and $N$ smallest values -- or, more precisely, into equally large sets $A$ and $B$ such that for any $a \in A$ and $b \in B$, $f_a \geq f_b$. These sets can then be subdivided into the elements starting in the lower-energy state ($A_0$ and $B_0$) and those starting in the higher-energy state ($A_1$ and $B_1$). $A_1$ and $B_0$ will have the same number of elements, so every element of $A_1$ can be paired with one unique element of $B_0$. If each of these pairs are exchanged, then all members of $A$ will be in the lower-energy state, and the system will be in a ground state, having released energy $\WG = ||A_1|| \varepsilon$. 

If those same pairs of cells are diffusively averaged rather than being exchanged, the released energy will be $(1/2) \WG$. This will often not be the optimal diffusive strategy, but it demonstrates that $\WD \geq (1/2) \WG$, which is enough to show (for the $2 \times N$ case) that $\WDmu / \WD \leq \WGmu / \WG$. 

However, this inequality does not always hold for systems with more than two allowed energy levels. Proving for a particular case that $\WDmu / \WD > \WGmu / \WG$ is often somewhat involved; given the other three available energies, it requires establishing an \textit{upper} bound for $\WD$. Consider the following configuration, with three energy levels and two allowed values for $\mu$:
\begin{align*}
\begin{array}{|c|c|c|}
\hline
X+1 & X & 0 \\ \hline
0 & 2 & 1 \\ \hline
\end{array} \; .
\end{align*}
Suppose the three columns correspond to energies $0$, $\varepsilon$, and $2 \varepsilon$, and consider the limit where $X \rightarrow \infty$. 

When $\mu$ conservation is enforced, the upper row is in its ground state, does not contribute to $\WGmu$ or $\WDmu$. The contribution from the lower row gives $\WGmu = 3 \varepsilon$. Applying the criteria for extremal sequences from Hay, Schiff, and Fisch\cite{Hay2015} for a three-level system, and checking all possible candidates, it is possible to show that the optimal sequence respecting $\mu$ conservation is 
\begin{align*}
\begin{array}{|c|c|c|}
\hline
X+1 & X & 0 \\ \hline
0 & 2 & 1 \\ \hline
\end{array} \rightarrow 
\begin{array}{|c|c|c|}
\hline
X+1 & X & 0 \\ \hline
1/2 & 2 & 1/2 \\ \hline
\end{array} \rightarrow\begin{array}{|c|c|c|}
\hline
X+1 & X & 0 \\ \hline
5/4 & 5/4 & 1/2 \\ \hline
\end{array} \; ,
\end{align*}
so that $\WDmu = (7/4) \varepsilon$. 

In the absence of $\mu$ conservation, the Gardner available energy is $\WG = (X + 1) \varepsilon$. In the limit where $X$ is very large, the only thing that will determine $\WD$ will be the fraction of $X$ that can be moved from the second column to the first. There is no strategy that can move more than half of the content of one cell to another using diffusive exchanges; in other words, $\lim_{X \rightarrow \infty} \WD = (X/2) \varepsilon$. Then, for this example, $\WDmu / \WD > \WGmu / \WG$. 

\section{Example:  Inhomogeneous Magnetic Field} \label{sec:inhomogeneousField}

In many scenarios, the kind of simple phase-space grid considered in the previous section may need to be modified. However, much of the intuition remains the same. 
One case that was discussed by Helander\cite{Helander2017ii} for the continuous restacking problem was a plasma in an inhomogeneous magnetic field, with conservation of the first adiabatic invariant $\mu = m v_\perp^2 / 2 B$. 
To illustrate the four free energies, consider for simplicity the case where the physical volume is divided into two halves:  one in which $B = B_0 = \text{const}$;  and another in which $B = B_1 = \text{const}$, with $B_1 > B_0$. 
Suppose the field is straight and its direction does not vary, and suppose the energy in the direction parallel to the field can be ignored. 
For simplicity, we consider a discrete version of this system.  The discrete version can be constructed by averaging the plasma distribution function over  finite regions of phase space, and then restricting the Gardner and diffusive exchange operations to act on these macroscopic regions. 

There are two major ways in which this scenario differs from those discussed in Section~\ref{sec:discreteModel}. First, for any given $\mu$, the volume of accessible phase space is now proportional to $B$. This can be shown by calculating the appropriate Jacobian determinant. Intuitively, it can be understood in terms of the geometry of phase space when $\mu$ is conserved. For a given $B$ and a given $\mu$ (or a given small range in $\mu$), the allowed region in phase space traces out a circle (or thin ring) in the $v_\perp = v_x \times v_y$ plane. If $\mu$ is held fixed and $B$ is changed from $B_0$ to $B_1$, then the region transforms to become a ring with a larger radius, with a correspondingly larger phase-space volume. The discrete system is composed of a series of boxes with equal phase-space volumes. Therefore, for any given $\mu$, there will be a larger number of phase-space boxes on the higher-field, higher-energy side of the system than on the lower-field, lower-energy side. 

\begin{figure}
	\includegraphics[width=\linewidth]{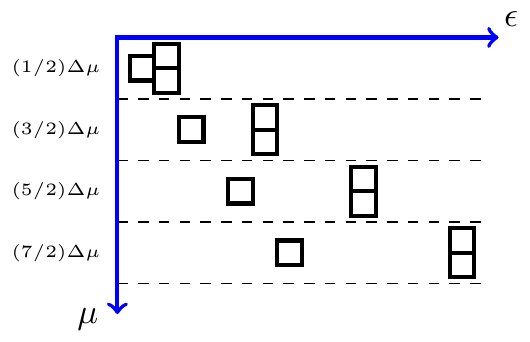}
	\caption{Schematic of the discrete restacking problem for a spatial region divided between an area with constant field $B_0$ and an area with constant field $B_1 = 2 B_0$. Dashed lines separate different values of $\mu$. For each value of $\mu$, the single box occupies the phase space on the low-field side and the pair of boxes occupy the phase space on the high-field side.} 
	\label{fig:newGrid}
\end{figure}

In addition, there is now a coupling between the invariant $\mu$ and the energy $\varepsilon$. 
For any given value of $B$, $\varepsilon \propto \mu$. 
Moreover, if a particle has energy $\varepsilon_0$ on the lower-field side, and if it is then moved to the higher-field side without changing $\mu$, it must then have energy $\varepsilon_1 = \mu (B_1 / B_0) \varepsilon_0$. When $\mu$ is conserved this is a two-energy-level system, but for different values of $\mu$ the difference between the two energy levels will be different. 
As a result, for this scenario,  analogous  to Figure~\ref{fig:grid}, Figure~\ref{fig:newGrid} represents the magnetic field constraint. 

This picture can be used to illustrate the behavior discussed by Helander. 
For example, if $\mu$ is conserved, then any distribution that is a function of $\mu$ alone is a ground state, and such states will have spatial densities that are proportional to $B$. 
This is in some sense counterintuitive, since for any given $\mu$ a higher field means a higher energy. 
However, the intuition can be restored with reference to Figure~\ref{fig:newGrid}. 
If $f$ is a function of $\mu$ alone, then each of the boxes in a given row must have the same population. 
It is then clear that distributions of this kind will give up no energy not only under $\mu$-conserving exchanges, but also under $\mu$-conserving diffusive exchanges. 

Moreover, it is clear why the total spatial density variations in such a state must be proportional to $B$: there are proportionally more boxes on the higher-field side than on the lower-field side for each choice of 
$\mu$, and therefore there must be a proportionally higher total population (in other words, the sum of the population over all $\mu$ for a given side of the system will be proportional to $B$). 
Helander also showed that $f = f(\mu)$ ground states will have temperatures proportional to $B(\bx)$. 
This can be understood in terms of Figure~\ref{fig:newGrid} in a similar way. 
Suppose $f = f(\mu)$. Then if the distribution function has some structure at an energy $\varepsilon$ on the low-field side, it must have the same structure at energy $(B_1/B_0) \varepsilon$ on the high-field side, since for a given $\mu$ the low-field and high-field regions of phase space have energies proportional to their local values of $B$. 
Given that the density of volumes is higher in the high field region proportionately to $(B_1/B_0)$, it follows that the pressure in high field regions is then  $(B_1/B_0)^2$ the pressure of the low field region. 

This is, of course, only one particular ground state, which occurs when all accessible states have equal occupation for each $\mu$.  
It is also possible to release free energy when the initial states are populated differently in such a way that there is population inversion, $(\partial f / \partial \epsilon)_\mu \ge 0$. 
In that case, for example with the higher B states populated equally but not the lower B state, representable by  $(0,1,1)$, then  the restacking solutions give one $\varepsilon$ free energy, whereas the diffusive solution allows only $(3/4) \varepsilon$ free energy (where $\varepsilon$ is the energy gap between the states). More generally, the difference between the diffusive and restacking solutions is similar to what was discussed in Section~\ref{sec:discreteModel}. However, the details do turn out somewhat differently. For instance, recall the first example from Section~\ref{sec:discreteModel}:
\begin{gather*}
\begin{array}{|c|c|}
\hline
0 & 1 \\ \hline
0 & 0 \\ \hline
\end{array} \; .
\end{gather*}
It is possible to construct a six-cell analog of this, with one populated cell on the high-field side and two possible choices of $\mu$, with a structure along the lines of the first two rows of Figure~\ref{fig:newGrid}. 
It might look something like this: 
\begin{align*}
\begin{array}{|c|}
\hline 0 \\ \hline
\end{array}
\begin{array}{|c|}
\hline 1 \\ 
\hline 0 \\ \hline
\end{array}& \\
&\begin{array}{|c|}
\hline 0 \\ \hline
\end{array}
\begin{array}{c}
~
\end{array}
\begin{array}{|c|}
\hline 0 \\
\hline 0 \\ \hline
\end{array}
\end{align*}
or this: 
\begin{align*}
\begin{array}{|c|}
\hline 0 \\ \hline
\end{array}
\begin{array}{|c|}
\hline 0 \\ 
\hline 0 \\ \hline
\end{array}& \\
&\begin{array}{|c|}
\hline 0 \\ \hline
\end{array}
\begin{array}{c}
~
\end{array}
\begin{array}{|c|}
\hline 1 \\
\hline 0 \\ \hline
\end{array} \; .
\end{align*}
But now the solutions will be different. For instance, it now makes a difference to all four available energies whether the populated cell was for the larger or smaller value of $\mu$, since these now have different energies and different gaps between their energies and the low-field states at the same $\mu$. 

\begin{figure}
	\includegraphics[width=\linewidth]{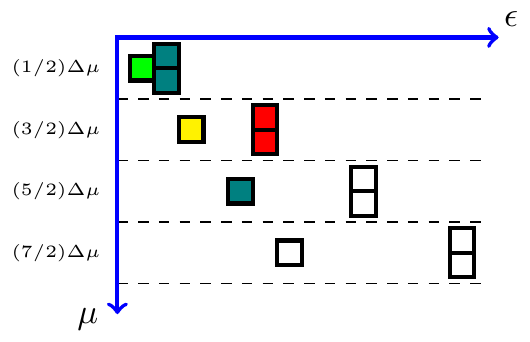}
	\caption{This version of Fig.~\ref{fig:newGrid} has cells marked with different colors in order to make the last example in Section~\ref{sec:inhomogeneousField} clearer.} 
	\label{fig:newGrid_marked}
\end{figure}

Now consider an example in which, for the second-lowest choice of $\mu$, the high-field states (marked red in Fig.~\ref{fig:newGrid_marked}) are occupied with populations normalized to 1 each, and the rest of phase space is unoccupied. 
This is analogous to the second example in Section~\ref{sec:discreteModel}. In this example, if each discrete box has a width $\Delta \mu$ in $\mu$ space, the bright green box has energy $(1/2) B_0 \Delta \mu$, the yellow box has energy $(3/2) B_0 \Delta \mu$, and each red box has energy $3 B_0 \Delta \mu$. 

If $\mu$ conservation is enforced, then the populations in the red boxes can only exchange with the box marked yellow in Fig.~\ref{fig:newGrid_marked}. The resulting diffusive free energy is $\WDmu = (9/8) B_0 \Delta \mu$. In the absence of a $\mu$ constraint, computing $\WD$ involves the red, yellow, green, and teal boxes. It is straightforward to find a bound on $\WD$ by considering only exchanges between the red, yellow, and bright green boxes: $\WD \geq (21/8) B_0 \Delta \mu$. 
Thus, the release of the constraint on the $\mu$ invariance makes substantially more energy available. 
This case provides some practical insight into $\alpha$-channeling; this will be discussed in Section~\ref{sec:discussion}. 

The system in this section has included only the component of the kinetic energy from motion perpendicular to the field. The qualitative behavior would be much the same if the parallel component were included. The phase space structure for any given $v_{||}$ would be identical to the one shown in Fig.~\ref{fig:newGrid}, but with an offset in energy depending on the value of $v_{||}$. However, there would be significant differences: the energy of a state would no longer be a function of $\mu$ and $B$ alone, so for any given $\mu$ a range of energies would be accessible on both the low- and high-field sides. 

\section{Summary and Discussion} \label{sec:discussion}

The key point of this paper is primarily one of  classification. There are in fact four free energies that are of interest in plasma:   the unconstrained restacking energy ($\WG$), 
the restacking energy with conservation laws ($\WGmu$),  the diffusive-exchange energy ($\WD$),  and the diffusive-exchange energy with conservation laws ($\WDmu$). 
The first three of these had been discussed previously in the literature.  
The last is first  described here. 

This classification is primarily of academic interest, at least insofar as  any detailed calculation of the energy is concerned. 
In practice, because accessing the full range of excitations is impractical, the full free energy will not be made available in any of of these categories.
Thus, for example, the precise Gardner restacking energy is not important as a quantity so much as it is important as a concept.  
As a concept, it exposes the fundamental physics of free energy under  phase space conservation.  
As a quantity, it does gives a useful measure of how much energy is {\it in principle} extractable, thus constructing a measure of success in extracting energy, or a measure of concern if the energy were lost to instabilities.
Similar arguments can be made for the worth of classifying the other free energies.

However, it is worth pointing out that, in recognizing the fourth free energy here, namely the diffusive-exchange energy with conservation laws ($\WDmu$), certain subtleties required quite careful attention in applying constraints to diffusive processes, as highlighted by the examples given.
For example, in the case of applying the $\mu$-conservation constraint in an inhomogeneous magnetic field $B$, the density of states needed to be recognized as proportional to $B$.  
This led to a very different, and somewhat non-intuitive, calculation  for the free energy under diffusion but respecting $\mu$-conservation. 

In fact, apart from its academic interest, this example may have an important application in $\alpha$-channeling.  
In the case of $\alpha$-channeling, $\alpha$-particles are diffused from the hot tokamak center by waves, which in doing so extract the $\alpha$-particle energy.\cite{Fisch1992}
However, generally one wave does not have all the necessary wave characteristics to accomplish this on its own.
It turns out that two waves, even if each is without necessary wave characteristics, together might be able to accomplish efficient energy extraction.\cite{Fisch1995}
In this regard, waves, such as the ion Bernstein wave, that break $\mu$ invariance\cite{Fisch1995ii} were proposed in combination with low-frequency waves that respect $\mu$ invariance to extract a substantial fraction of the $\alpha$-particle energy.\cite{Herrmann1997}

The strategy that achieved the most energy extraction (of about 61\%) featured diffusion by the ion Bernstein wave confined to an outer region, in keeping with a realistic spatial deployment of  this wave.
In other regions, the low frequency waves dominated the diffusion.  
This strategy, while giving quite favorable results, was by no means necessarily optimal, since the space of wave parameters to search is extremely large.

Note that waves that do not break the $\mu$ invariant can diffuse $\alpha$-particles from the tokamak center, where they are born energetic, to the low-field side of the tokamak. 
Thus, in conserving $\mu$, the $\alpha$-particles advantageously lose perpendicular energy. 
However, as discussed in Section IV, the density of states at any given $\mu$ is higher in the tokamak center where the magnetic field is larger than on the periphery where the magnetic field is smaller. This statement pertains even though the phase space on tokamaks is six-dimensional (since it includes the parallel velocity) rather than the four-dimensional phase space considered for simplicity in Section~\ref{sec:inhomogeneousField}. In either case, the diffusion to lower magnetic fields, when respecting the $\mu$ invariance, whether in four or six dimensions, will be to states less dense, and therefore the energy extraction will be less efficient than when diffusion occurs to states equally dense. 
In the last example considered in the previous section, where the magnetic fields on either side of the system differed by a factor of two and the system began with a population of particles on the high-field side, the extractable energy under diffusion and respecting $\mu$ invariance was less than half of the extractable energy when $\mu$ invariance was not respected. 

Much of the previous work on two-wave $\alpha$-channeling has rested on the intuition that a high-frequency, non-$\mu$-conserving wave could help to extract the energy from $\alpha$-particles while a low-frequency, $\mu$-conserving wave could accomplish spatial diffusion. The work presented here suggests a second motivation for combining such waves: that breaking $\mu$ conservation may also allow diffusive processes to more efficiently extract energy. 
In other words, there emerges the interesting suggestion  that a greater amount of energy may become available by arranging for breaking the  $\mu$ invariance even in regions where the low-frequency waves dominate the diffusion.  
Since the parameter space of wave possibilities is quite large, such an intuition may prove valuable in optimizing the  $\alpha$-channeling effect.
At present, this is a speculation, not a proved result. 
A rigorous proof would require a much more specialized analysis of the initial conditions and phase-space geometry associated with $\alpha$-channeling, which goes beyond the intended scope of this paper. 
In addition, of course, it would be important to show that such a scheme can work for specific wave implementations.

The above example shows how consideration of the new, fourth free energy identified here leads to useful, possibly practical, physical intuitions regarding extractable energy under wave diffusion by different waves. 
However, in the end, the major interest of this classification remains an academic understanding for the fundamental ways in which energy may be released in different plasma systems.

\begin{acknowledgements}
This work was supported by  US DOE grants  DE-AC02-09CH11466 and DE-SC0016072. 
One of the authors (PH) would like to acknowledge the splendid hospitality of the Princeton Plasma Physics Laboratory. 
\end{acknowledgements}

\section*{Data Availability Statement}

Data sharing is not applicable to this article as no new data were created or analyzed in this study.

\bibliographystyle{apsrev4-1} 
	\bibliography{../Master.bib}

\providecommand{\noopsort}[1]{}\providecommand{\singleletter}[1]{#1}%
\begin{thebibliography}{16}%
\makeatletter
\providecommand \@ifxundefined [1]{%
 \@ifx{#1\undefined}
}%
\providecommand \@ifnum [1]{%
 \ifnum #1\expandafter \@firstoftwo
 \else \expandafter \@secondoftwo
 \fi
}%
\providecommand \@ifx [1]{%
 \ifx #1\expandafter \@firstoftwo
 \else \expandafter \@secondoftwo
 \fi
}%
\providecommand \natexlab [1]{#1}%
\providecommand \enquote  [1]{``#1''}%
\providecommand \bibnamefont  [1]{#1}%
\providecommand \bibfnamefont [1]{#1}%
\providecommand \citenamefont [1]{#1}%
\providecommand \href@noop [0]{\@secondoftwo}%
\providecommand \href [0]{\begingroup \@sanitize@url \@href}%
\providecommand \@href[1]{\@@startlink{#1}\@@href}%
\providecommand \@@href[1]{\endgroup#1\@@endlink}%
\providecommand \@sanitize@url [0]{\catcode `\\12\catcode `\$12\catcode
  `\&12\catcode `\#12\catcode `\^12\catcode `\_12\catcode `\%12\relax}%
\providecommand \@@startlink[1]{}%
\providecommand \@@endlink[0]{}%
\providecommand \url  [0]{\begingroup\@sanitize@url \@url }%
\providecommand \@url [1]{\endgroup\@href {#1}{\urlprefix }}%
\providecommand \urlprefix  [0]{URL }%
\providecommand \Eprint [0]{\href }%
\providecommand \doibase [0]{http://dx.doi.org/}%
\providecommand \selectlanguage [0]{\@gobble}%
\providecommand \bibinfo  [0]{\@secondoftwo}%
\providecommand \bibfield  [0]{\@secondoftwo}%
\providecommand \translation [1]{[#1]}%
\providecommand \BibitemOpen [0]{}%
\providecommand \bibitemStop [0]{}%
\providecommand \bibitemNoStop [0]{.\EOS\space}%
\providecommand \EOS [0]{\spacefactor3000\relax}%
\providecommand \BibitemShut  [1]{\csname bibitem#1\endcsname}%
\let\auto@bib@innerbib\@empty
%</preamble>
\bibitem [{\citenamefont {Gardner}(1963)}]{Gardner1963}%
  \BibitemOpen
  \bibfield  {author} {\bibinfo {author} {\bibfnamefont {C.~S.}\ \bibnamefont
  {Gardner}},\ }\href {\doibase 10.1063/1.1706823} {\bibfield  {journal}
  {\bibinfo  {journal} {Phys. Fluids}\ }\textbf {\bibinfo {volume} {6}},\
  \bibinfo {pages} {839} (\bibinfo {year} {1963})}\BibitemShut {NoStop}%
\bibitem [{\citenamefont {Dodin}\ and\ \citenamefont
  {Fisch}(2005)}]{Dodin2005}%
  \BibitemOpen
  \bibfield  {author} {\bibinfo {author} {\bibfnamefont {I.~Y.}\ \bibnamefont
  {Dodin}}\ and\ \bibinfo {author} {\bibfnamefont {N.~J.}\ \bibnamefont
  {Fisch}},\ }\href {\doibase 10.1016/j.physleta.2005.04.078} {\bibfield
  {journal} {\bibinfo  {journal} {Phys. Lett. A}\ }\textbf {\bibinfo {volume}
  {341}},\ \bibinfo {pages} {187} (\bibinfo {year} {2005})}\BibitemShut
  {NoStop}%
\bibitem [{\citenamefont {Helander}(2017)}]{Helander2017ii}%
  \BibitemOpen
  \bibfield  {author} {\bibinfo {author} {\bibfnamefont {P.}~\bibnamefont
  {Helander}},\ }\href {\doibase 10.1017/S0022377817000496} {\bibfield
  {journal} {\bibinfo  {journal} {J. Plasma Phys.}\ }\textbf {\bibinfo {volume}
  {83}},\ \bibinfo {pages} {715830401} (\bibinfo {year} {2017})}\BibitemShut
  {NoStop}%
\bibitem [{\citenamefont {Helander}(2020)}]{Helander2020}%
  \BibitemOpen
  \bibfield  {author} {\bibinfo {author} {\bibfnamefont {P.}~\bibnamefont
  {Helander}},\ }\href {\doibase 10.1017/S0022377820000057} {\bibfield
  {journal} {\bibinfo  {journal} {J. Plasma Phys.}\ }\textbf {\bibinfo {volume}
  {86}},\ \bibinfo {pages} {905860201} (\bibinfo {year} {2020})}\BibitemShut
  {NoStop}%
\bibitem [{\citenamefont {Kennel}\ and\ \citenamefont
  {Engelmann}(1966)}]{Kennel1966}%
  \BibitemOpen
  \bibfield  {author} {\bibinfo {author} {\bibfnamefont {C.~F.}\ \bibnamefont
  {Kennel}}\ and\ \bibinfo {author} {\bibfnamefont {F.}~\bibnamefont
  {Engelmann}},\ }\href {\doibase 10.1063/1.1761629} {\bibfield  {journal}
  {\bibinfo  {journal} {Phys. Fluids}\ }\textbf {\bibinfo {volume} {9}},\
  \bibinfo {pages} {2377} (\bibinfo {year} {1966})}\BibitemShut {NoStop}%
\bibitem [{\citenamefont {Fisch}\ and\ \citenamefont {Rax}(1992)}]{Fisch1992}%
  \BibitemOpen
  \bibfield  {author} {\bibinfo {author} {\bibfnamefont {N.~J.}\ \bibnamefont
  {Fisch}}\ and\ \bibinfo {author} {\bibfnamefont {J.-M.}\ \bibnamefont
  {Rax}},\ }\href {\doibase 10.1103/PhysRevLett.69.612} {\bibfield  {journal}
  {\bibinfo  {journal} {Phys. Rev. Lett.}\ }\textbf {\bibinfo {volume} {69}},\
  \bibinfo {pages} {612} (\bibinfo {year} {1992})}\BibitemShut {NoStop}%
\bibitem [{\citenamefont {Fisch}\ and\ \citenamefont {Rax}(1993)}]{Fisch1993}%
  \BibitemOpen
  \bibfield  {author} {\bibinfo {author} {\bibfnamefont {N.~J.}\ \bibnamefont
  {Fisch}}\ and\ \bibinfo {author} {\bibfnamefont {J.-M.}\ \bibnamefont
  {Rax}},\ }\href {\doibase 10.1063/1.860809} {\bibfield  {journal} {\bibinfo
  {journal} {Phys. Fluids B}\ }\textbf {\bibinfo {volume} {5}},\ \bibinfo
  {pages} {1754} (\bibinfo {year} {1993})}\BibitemShut {NoStop}%
\bibitem [{\citenamefont {Dalton}(1920)}]{Dalton1920}%
  \BibitemOpen
  \bibfield  {author} {\bibinfo {author} {\bibfnamefont {H.}~\bibnamefont
  {Dalton}},\ }\href {\doibase 10.2307/2223525} {\bibfield  {journal} {\bibinfo
   {journal} {Econ. J.}\ }\textbf {\bibinfo {volume} {30}},\ \bibinfo {pages}
  {348} (\bibinfo {year} {1920})}\BibitemShut {NoStop}%
\bibitem [{\citenamefont {Thon}\ and\ \citenamefont
  {Wallace}(2004)}]{Thon2004}%
  \BibitemOpen
  \bibfield  {author} {\bibinfo {author} {\bibfnamefont {D.}~\bibnamefont
  {Thon}}\ and\ \bibinfo {author} {\bibfnamefont {S.~W.}\ \bibnamefont
  {Wallace}},\ }\href {\doibase 10.1007/s00355-003-0226-x} {\bibfield
  {journal} {\bibinfo  {journal} {Soc. Choice Welfare}\ }\textbf {\bibinfo
  {volume} {22}},\ \bibinfo {pages} {447} (\bibinfo {year} {2004})}\BibitemShut
  {NoStop}%
\bibitem [{\citenamefont {Horn}(1964)}]{Horn1964}%
  \BibitemOpen
  \bibfield  {author} {\bibinfo {author} {\bibfnamefont {F.}~\bibnamefont
  {Horn}},\ }\enquote {\bibinfo {title} {Attainable and non-attainable regions
  in chemical reaction techniques},}\ in\ \href@noop {} {\emph {\bibinfo
  {booktitle} {Proceedings of the 3rd European Symposium on Chemical Reaction
  Engineering}}}\ (\bibinfo  {publisher} {Pergamon},\ \bibinfo {year} {1964})\
  pp.\ \bibinfo {pages} {1--10}\BibitemShut {NoStop}%
\bibitem [{\citenamefont {Zylka}(1985)}]{Zylka1985}%
  \BibitemOpen
  \bibfield  {author} {\bibinfo {author} {\bibfnamefont {C.}~\bibnamefont
  {Zylka}},\ }\href {\doibase 10.1007/BF00529057} {\bibfield  {journal}
  {\bibinfo  {journal} {Theor. Chim. Acta}\ }\textbf {\bibinfo {volume} {68}},\
  \bibinfo {pages} {363} (\bibinfo {year} {1985})}\BibitemShut {NoStop}%
\bibitem [{\citenamefont {Hay}\ \emph {et~al.}(2015)\citenamefont {Hay},
  \citenamefont {Schiff},\ and\ \citenamefont {Fisch}}]{Hay2015}%
  \BibitemOpen
  \bibfield  {author} {\bibinfo {author} {\bibfnamefont {M.~J.}\ \bibnamefont
  {Hay}}, \bibinfo {author} {\bibfnamefont {J.}~\bibnamefont {Schiff}}, \ and\
  \bibinfo {author} {\bibfnamefont {N.~J.}\ \bibnamefont {Fisch}},\ }\href
  {\doibase 10.1063/1.4933018} {\bibfield  {journal} {\bibinfo  {journal}
  {Phys. Plasmas}\ }\textbf {\bibinfo {volume} {22}},\ \bibinfo {pages}
  {102108} (\bibinfo {year} {2015})}\BibitemShut {NoStop}%
\bibitem [{\citenamefont {Hay}\ \emph {et~al.}(2017)\citenamefont {Hay},
  \citenamefont {Schiff},\ and\ \citenamefont {Fisch}}]{Hay2017}%
  \BibitemOpen
  \bibfield  {author} {\bibinfo {author} {\bibfnamefont {M.~J.}\ \bibnamefont
  {Hay}}, \bibinfo {author} {\bibfnamefont {J.}~\bibnamefont {Schiff}}, \ and\
  \bibinfo {author} {\bibfnamefont {N.~J.}\ \bibnamefont {Fisch}},\ }\href
  {\doibase 10.1016/j.physa.2017.01.038} {\bibfield  {journal} {\bibinfo
  {journal} {Physica A}\ }\textbf {\bibinfo {volume} {473}},\ \bibinfo {pages}
  {225} (\bibinfo {year} {2017})}\BibitemShut {NoStop}%
\bibitem [{\citenamefont {Fisch}\ and\ \citenamefont
  {Herrmann}(1995)}]{Fisch1995}%
  \BibitemOpen
  \bibfield  {author} {\bibinfo {author} {\bibfnamefont {N.~J.}\ \bibnamefont
  {Fisch}}\ and\ \bibinfo {author} {\bibfnamefont {M.~C.}\ \bibnamefont
  {Herrmann}},\ }\href {\doibase 10.1088/0029-5515/35/12/I40} {\bibfield
  {journal} {\bibinfo  {journal} {Nucl. Fusion}\ }\textbf {\bibinfo {volume}
  {35}},\ \bibinfo {pages} {1753} (\bibinfo {year} {1995})}\BibitemShut
  {NoStop}%
\bibitem [{\citenamefont {Fisch}(1995)}]{Fisch1995ii}%
  \BibitemOpen
  \bibfield  {author} {\bibinfo {author} {\bibfnamefont {N.~J.}\ \bibnamefont
  {Fisch}},\ }\href {\doibase 10.1063/1.871454} {\bibfield  {journal} {\bibinfo
   {journal} {Phys. Plasmas}\ }\textbf {\bibinfo {volume} {2}},\ \bibinfo
  {pages} {2375} (\bibinfo {year} {1995})}\BibitemShut {NoStop}%
\bibitem [{\citenamefont {Herrmann}\ and\ \citenamefont
  {Fisch}(1997)}]{Herrmann1997}%
  \BibitemOpen
  \bibfield  {author} {\bibinfo {author} {\bibfnamefont {M.~C.}\ \bibnamefont
  {Herrmann}}\ and\ \bibinfo {author} {\bibfnamefont {N.~J.}\ \bibnamefont
  {Fisch}},\ }\href {\doibase 10.1103/PhysRevLett.79.1495} {\bibfield
  {journal} {\bibinfo  {journal} {Phys. Rev. Lett.}\ }\textbf {\bibinfo
  {volume} {79}},\ \bibinfo {pages} {1495} (\bibinfo {year}
  {1997})}\BibitemShut {NoStop}%
\end{thebibliography}%
%	\bibliography{Master}
	
\end{document}